\documentclass[aps,prl,amssymb,showpacs,twocolumn]{revtex4}
\usepackage{graphicx}
\usepackage{bm}

\begin{document}

\title{Predicting crystal structures: the Parrinello-Rahman method revisited}
\author{R. Marto\v{n}\'ak} \altaffiliation[Permanent address: ]{Department
  of Physics, Faculty of Electrical Engineering and Information Technology,
  Slovak University of Technology, Ilkovi\v{c}ova 3, 812 19 Bratislava,
  Slovakia}
\affiliation{Swiss Center for Scientific Computing, Via Cantonale, CH-6928
  Manno, Switzerland and ETH Zurich, Physical Chemistry, Hoenggerberg,
  CH-8093 Zurich, Switzerland}
\author{A. Laio} \affiliation{Swiss Center for Scientific Computing, Via
  Cantonale, CH-6928 Manno, Switzerland and ETH Zurich, Physical Chemistry,
  Hoenggerberg, CH-8093 Zurich, Switzerland}
\author{M. Parrinello} \affiliation{Swiss Center for Scientific Computing,
  Via Cantonale, CH-6928 Manno, Switzerland and ETH Zurich, Physical
  Chemistry, Hoenggerberg, CH-8093 Zurich, Switzerland}
 \date{\today}

\begin{abstract}
  By suitably adapting a recent approach [A. Laio and M. Parrinello, PNAS,
  \textbf{99}, 12562 (2002)] we develop a powerful molecular dynamics
  method for the study of pressure-induced structural transformations. We
  use the edges of the simulation cell as collective variables. In the
  space of these variables we define a metadynamics that drives the system
  away from the local minimum towards a new crystal structure. In contrast
  to the Parrinello-Rahman method our approach shows no hysteresis and
  crystal structure transformations can occur at the equilibrium pressure.
  We illustrate the power of the method by studying the pressure-induced
  diamond to simple hexagonal phase transition in a model of silicon.
\end{abstract}

\maketitle

Predicting equilibrium crystal structures at a given pressure and
temperature is an important problem in fields of science as different as
solid state physics, geophysics, planetary physics, materials science,
polymer science, chemistry, etc. Upon increasing external pressure crystals
usually undergo structural phase transitions. Often, the final structure is
unknown and simulations can be very useful in identifying possible
candidates. This task represents a challenge for computational physics.
Great progress was achieved with the introduction of constant-pressure
molecular dynamics (MD) \cite{andersen} and in particular the
Parrinello-Rahman method \cite{pr} where the box is allowed to change its
shape in order to comply with a new structure. The Parrinello-Rahman method
is now described in textbooks and widely used also in different
variants\cite{klein,ray,hoover,wentz,lill,melchionna,martyna,martins}.
However, structural transitions are often first order. Since for crystal
simulations periodic boundary conditions are commonly used, heterogeneous
nucleation is suppressed and the system has to cross a significant barrier
in order to transform from one structure to another. As a consequence large
hysteretic effects are observed within the above approaches. In order to
observe the transition within the accessible simulation time\cite{kaxiras}
one often has to overpressurize the system close to the point of mechanical
instability.  Under such conditions one or more intermediate phases may be
skipped\cite{martins,marco}, which reduces the predictive power of the
method.

In the work of Parrinello and Rahman\cite{pr} it was realized that in an MD
simulation of a crystal phase transition it is necessary to treat the MD
supercell edges $\vec{a},\vec{b},\vec{c}$ as dynamical variables.  These
variables were arranged to form a matrix
$\mathbf{h}=(\vec{a},\vec{b},\vec{c})$ and extending the work of Andersen
\cite{andersen} a Lagrangian was introduced that coupled the $\mathbf{h}$
degrees of freedom with the microscopic motion of the atoms under condition
of constant pressure. Due to the time-scale problem mentioned earlier this
approach tends to be ineffective at pressures close to the critical
transition pressure. As the origin of the problem is the lack of efficiency
of standard molecular dynamics in crossing high barriers, we adopt here a
conceptually different strategy. Since our aim is to simulate a phase
transition at a pressure $P$ and a temperature $T$ we consider the Gibbs
potential $\mathcal{G}(\mathbf{h})=\mathcal{F}(\mathbf{h})+PV$ as a
function of $\mathbf{h}$ where $\mathcal{F}(\mathbf{h})$ is the Helmholtz
free energy of the system at fixed box and $V=\det(\mathbf{h})$ is the
volume of the box. We assume, following Nos\'e and Klein \cite{klein}, that
the matrix $\mathbf{h}$ is symmetric in order to eliminate rotations of the
supercell. This reduces the number of collective variables to 6. These 6
independent components of $\mathbf{h}$ now represent collective
coordinates, or order parameters, which distinguish the different minima of
$\mathcal{G}$.  We note that the derivative
\begin{equation}
 -\frac{\partial \mathcal{G}}{\partial h_{ij}} =  V \left\{\left[
     (\mathbf{p} - P)\mathbf{h}^{-1}\right]_{ij}  
+ \left[ (\mathbf{p} - P)\mathbf{h}^{-1}\right]_{ji} \right\} \left(1 -
\frac{1}{2}\delta_{ij} \right) \label{eq:1} 
\end{equation}
where $\mathbf{p}$ is the internal pressure tensor, can be easily evaluated
from microscopic MD or Monte Carlo runs at constant $\mathbf{h}$ by
averaging the microscopic virial tensor. Making use of Eq.(\ref{eq:1}) we
now construct an algorithm, based on the recently introduced method of
Ref.\cite{hills}, that is able to explore the surface
$\mathcal{G}(\mathbf{h})$ efficiently and in particular can identify the
local minima which correspond to stable or metastable crystal structures at
a given pressure $P$. The method\cite{hills} has been shown to be able to
dramatically speed up the simulation of activated processes and is
therefore well suited for simulating first-order phase transitions.  We
describe here the basic ideas and refer for more details to the original
paper.

Following Ref.\cite{hills}, the collective variables that are now arranged
to form a 6-dimensional vector $\mathbf{h} = (h_{11}, h_{22}, h_{33},
h_{12}, h_{13}, h_{23})$ are evolved according to a steepest-descent-like
discrete evolution with a stepping parameter $\delta h$
(\emph{metadynamics})
\begin{equation}
\mathbf{h}^{t+1}=\mathbf{h}^{t}+\delta h \frac{\bm{\phi}^{t}}{\left|
    \bm{\phi}^{t}\right|  } \; .
\label{eq:2}
\end{equation}
Here, the driving force $\bm{\phi}^{t} = - \frac{\partial G^{t} }{\partial
  \mathbf{h}}$ is derived from a history-dependent Gibbs potential
$\mathcal{G}^{t}$ where a Gaussian has been added to
$\mathcal{G}(\mathbf{h})$ at every point $\mathbf{h}^{t^{\prime }}$ already
visited in order to discourage it from being visited again. Hence we have
\begin{equation}
  \mathcal{G}^{t}(\mathbf{h}) = \mathcal{G}(\mathbf{h}) + \sum_{t^{\prime }
    < t} W 
  e^{-\frac{|\mathbf{h}-\mathbf{h}^{t^{\prime }}|^{2}}{2\delta h^{2}}} 
\label{eq:3} 
\end{equation}
and the force $\bm{\phi}^{t}$ is therefore a sum of a thermodynamical
driving force $\mathbf{F} = - \frac{\partial \mathcal{G} }{\partial
  \mathbf{h}}$ and the term $\mathbf{F_{g}}$ coming from a potential
constructed as a superposition of Gaussians.  As time proceeds the
history-dependent term in Eq.(\ref{eq:3}) fills the initial well of the
free-energy surface and the system is driven along the lowest free energy
pathway out of the local minimum. The passage through the transition state
can be detected by monitoring the relative orientation of the forces
$\mathbf{F}$ and $\mathbf{F_{g}}$. While a well is being filled these two
forces approximately balance each other, $\mathbf{F} + \mathbf{F_{g}}
\approx 0$, and the two vectors have roughly opposite directions. After
crossing the saddle point this is no longer true and $\mathbf{F}$ and
$\mathbf{F_{g}}$ become almost parallel and oriented along the eigenvector
corresponding to the negative eigenvalue of the Hessian matrix.  The
indicator $\mathbf{F} \cdot \mathbf{F_{g}}/(|\mathbf{F}| |\mathbf{F_{g}}|)$
develops a sharp spike which can be used to signal the transition from one
basin to the other.

The choice of the parameters $W$ and $\delta h$ depends on the
$\mathcal{G}(\mathbf{h})$ landscape. In order to achieve the necessary
energy resolution, $W$ should be chosen as a fraction of the relevant
energy barriers. The parameter $\delta h$ on the other hand determines the
resolution in $\mathbf{h}$. However, a very small value of $\delta h$ is
not to be recommended. In fact a small $\delta h$ requires longer runs.
Furthermore for an optimal filling the curvature of the Gaussians should be
smaller than that of the well. This leads to the condition
$\frac{W}{{\delta h}^2} \leq K$ where $K$ is the smallest eigenvalue of the
$\mathcal{G}(\mathbf{h})$ Hessian at the minimum $\mathbf{h_0}$. For a
cubic system we can estimate $K$ from the approximate expansion of
$\mathcal{G}(\mathbf{h})$ around $\mathbf{h_0}$
\begin{equation}
\mathcal{G}(\Delta \mathbf{h}) \approx \mathcal{G}(\mathbf{h_0}) +
\frac{1}{2} V c (\frac{\Delta \mathbf{h}}{L})^2 
\end{equation} 
where $L$ is the cell edge and $c$ is of the order of magnitude of the
elastic constants. This leads to the estimate $K \approx L c$ and to the
condition $\frac{W}{{\delta h}^2} \lesssim L c$. A more general discussion
of the choice of $W$ and $\delta h$ can be found in Ref.\cite{hills}.

In practice the metadynamics simulation proceeds as follows. We start from
an equilibrated value of $\mathbf{h}$ at a given pressure $P$ and
temperature $T$ and evaluate the pressure tensor $\mathbf{p}$ in a constant
$\mathbf{h}$ MD run long enough to allow relaxation to equilibrium and
sufficient averaging of $\mathbf{p}$.  The $\mathbf{h}$ is updated using
the forces (\ref{eq:1}) and metadynamics equations (\ref{eq:2},\ref{eq:3})
to a new value $\mathbf{h}^{'}$. After the box is modified the particle
positions are rescaled in order to fit into the new box using the relation
$\vec{r}^{'} = \mathbf{h}^{'} \mathbf{h}^{-1} \vec{r}$. As the initial free
energy well is gradually filled the box undergoes a set of deformations
until a transition state is reached and the system enters into the basin of
attraction of a new state.  In order to characterize the new phase it is
often useful at this stage to switch off the Gaussian term, so that the
metadynamics becomes purely steepest descent-like and drives the system
towards the equilibrium state for the new structure. In this equilibrium
state the pressure will be equal to $P$.  However, during the metadynamics
the pressure tensor can become anisotropic and the internal pressure may be
different from $P$.  Once the new structure is characterized one can switch
the Gaussians on again, thus filling the new minimum, and move to other
minima, if available.  The metadynamics is capable of reconstructing the
free energy profile\cite{hills}, since the sum of the Gaussians in
Eq.(\ref{eq:3}) converges to $- \mathcal{G}(\mathbf{h})$ up to an additive
constant, if $W$ and $\delta h$ are properly chosen. This will not be used
here since once the structures are known it is relatively straightforward
to calculate their free energy\cite{frenkel}.  We emphasize, however, that
free-energy calculations alone do not provide an alternative to our method
since they assume knowledge of the final crystal structure.

We have tested our method on several model Hamiltonians. Here we report
only a simulation of a tight-binding model of Si\cite{lenosky} at a
pressure very close to the theoretical transition pressure.  This
tight-binding parametrization captures some of the main features of the Si
phase diagram and provides a convenient test model. In the following we
shall use a supercell of 216 atoms and only the $\Gamma $ point of the BZ.
The $T=0$ phase diagram for this model system can be found by performing
energy versus volume calculations in the three relevant structures, namely
the $P=0$ equilibrium diamond structure and the two high-pressure phases,
$\beta$-tin and simple hexagonal (SH). The latter two are almost degenerate
in energy, $\beta$-tin being only metastable. A common tangent construction
gives a critical pressure of 15.5 GPa for the transition from the diamond
to the SH phase.  Applying the Parrinello-Rahman method to the same model
and system size, the transition from diamond to the SH phase is found to
occur at 44 GPa \cite{jahnatek}, which corresponds to an overpressurization
of almost a factor of 3. Here we show instead that with our new method the
transition can be observed with affordable computational effort at $T=300$
K and $P=16$ GPa, i.e. very close to the critical pressure.

A metadynamics simulation was run with the parameters $W = 8.6$ eV and
$\delta h =1$ {\AA} which are compatible with the guidelines given earlier,
taking into account that $L \approx$ 15 {\AA} and a typical Si elastic
constant value is $c \approx$ 100 GPa. We have preferred a relatively large
value for $\delta h$ in order to enhance volume fluctuations and thus to
favour the change of volume which accompanies the pressure-induced
transformation of diamond Si. This choice is also instrumental in avoiding
that the system makes a fake transition to the same crystal structure,
since such a transition obviously conserves volume. The origin of these
fake transformations is to be found in the fact that a given crystal
structure can correspond to different values of $\mathbf{h}$. This problem,
which is a consequence of what is known as modular invariance \cite{wentz},
can be fixed in many different ways. However, in view of the simple
solution found here we have not pursued these alternatives, which will be
discussed elsewhere.

At each metadynamics step we equilibrated the system for 1 ps and averaged
the pressure tensor for another ps; temperature was controlled by
Nos\'{e}-Hoover chains\cite{nose}. The history of the run is shown in
Fig.1. The indicator $\mathbf{F} \cdot \mathbf{F_{g}}/(|\mathbf{F}|
|\mathbf{F_{g}}|)$ (Fig.1(e)) clearly shows that at metastep 35 there is a
phase transition (see also Fig.1(d) and Fig.2(a)-(d)). Consequently after
this step we switched off the Gaussians to let the system evolve towards
the new structure in a steepest-descent-like manner till metastep 50 (see
Fig.2(e)-(h)). This final structure was then evolved for 5 ps with a
Parrinello-Rahman simulation.  During this time very little relaxation of
the cell parameters took place, which confirms that we have reached a
minimum of $\mathcal{G}(\mathbf{h})$.  A visual inspection of the final
structure (Fig.2(h)) as well as an analysis of the diffraction peaks showed
that the system had made a transition to the SH structure, whose parameters
are $a=2.61$ {\AA} and $c=2.48$ {\AA}.  We also calculated the atomic
volume, shown in Fig. 1(c).  After step 34 we observe a pronounced drop,
agreeing well with the results of the $T=0$ calculation which predicts a
change from 17.5 {\AA}$^3$ per atom in the diamond to 14.8 {\AA}$^3$ per
atom in the SH phase.

Another run was carried out in which after metastep 35 the Gaussians were
kept switched on.  This metadynamics led to a series of transitions between
the stable SH and metastable $\beta$-tin structures with different
orientations.  We also performed a simulation of decompression of the SH
structure at $p = 5$ GPa which after 48 metadynamics steps transformed into
a tetrahedrally-coordinated amorphous structure.  Experimentally Si upon
decompression also does not come back from the $\beta$-tin to the diamond
structure but transforms to a series of metastable structures \cite{crain}.
These two examples demonstrate that the method is able to find also
metastable crystalline and amorphous phases.

This calculation shows that the method overcomes many of the limitations of
the previous approaches. It must be stressed that, in contrast to previous
work\cite{andersen,pr,klein,ray,hoover,wentz,lill,melchionna,martyna,martins},
it is not a constant-pressure simulation method but rather a method for
exploring the dependence of the Gibbs free energy on the $\mathbf{h}$
variables. The use of the history-dependent metadynamics allows large
energy barriers to be overcome in a short time and makes this approach very
efficient.  If other internal slow degrees of freedom besides the
$\mathbf{h}$ variables are present, as in molecular crystals, these can
naturally be added and taken into account within the general
scheme\cite{hills}. Metadynamics is in principle able to visit any
crystalline structure that is at least metastable at a given pressure;
however, use of pressures considerably different from the critical one may
result in the need for a longer time in order to escape from the initial
minimum.  In the example presented here the total aggregated simulation
time is about 100 ps. This makes the method suitable also for an ab-initio
MD simulation of systems containing about 100 atoms.  Another advantage is
that ab-initio constant-volume codes can be used, avoiding the need to use
expensive tricks to deal with the Pulay correction\cite{marco}.  In
conclusion it can be confidently stated that this new method, with its
ability to induce structural transitions at equilibrium conditions, can
substantially improve the predictive power of solid-state simulations.

We would like to acknowledge the help of M. Jahn\'atek in the
implementation of the tight-binding method as well as stimulating
discussions with M. Bernasconi.

\begin{figure}[h]
\includegraphics*[width=8.6cm]{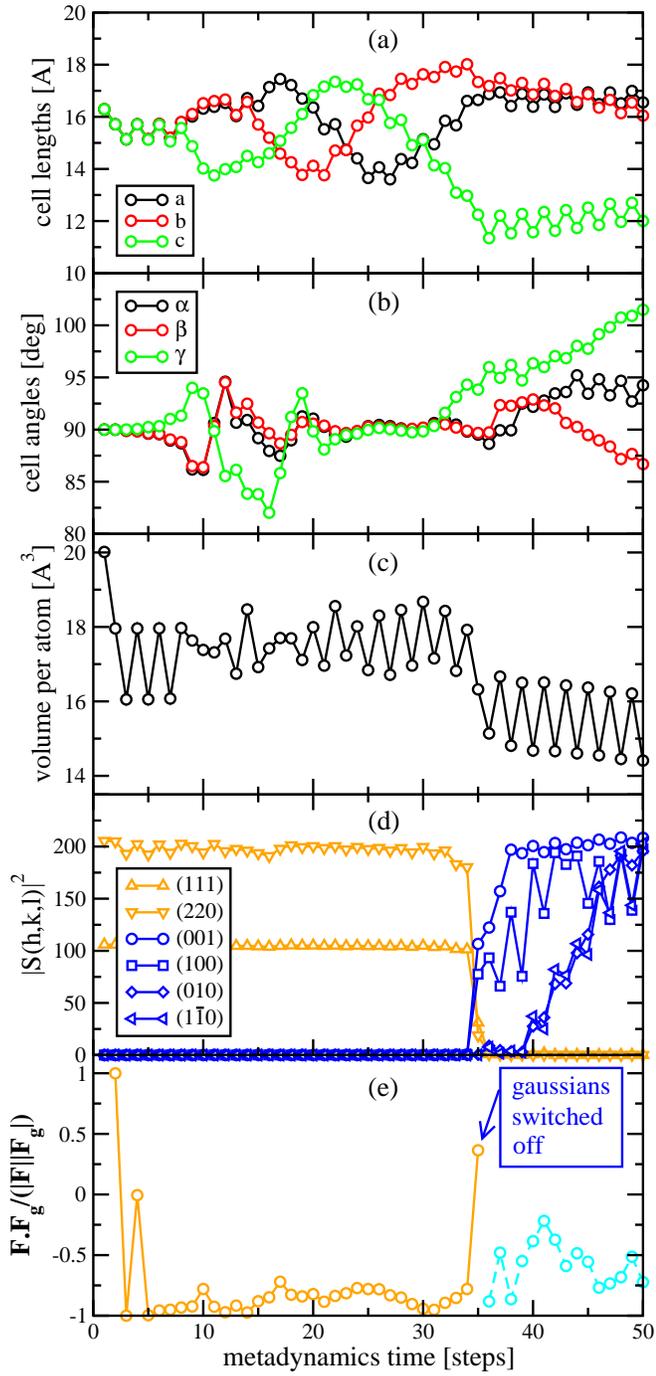}
\caption{Evolution of cell lengths (a), cell angles (b), atomic volume (c),
  selected peaks of the diamond (averaged over all equivalent directions,
  orange) and SH (blue) structure factor (d) and relative orientation of
  forces $\mathbf{F}$ and $\mathbf{F}_g$ (e) during the metadynamics. Note
  the structural transition at step 35. The Gaussian term in
  Eqn.(\ref{eq:3}) is switched off after step 35. The light blue curve in
  (e) shows the continuation of the orange run in the mode in which the
  Gaussians are added at every metadynamics step. }
\label{transition}
\end{figure}

\begin{figure}[h]
\includegraphics*[width=18cm]{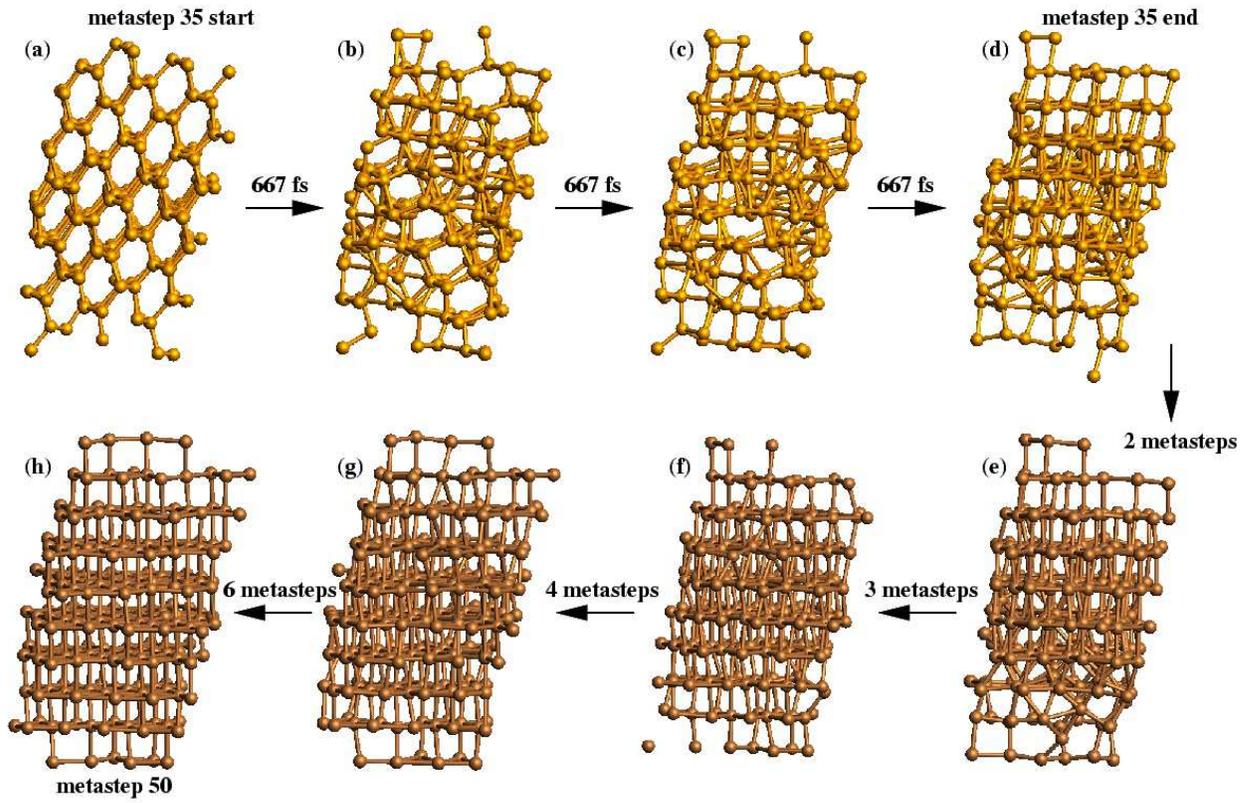}
\caption{(a) - (d) Evolution of atomic configurations during 2 ps of
  microscopic dynamics (at intervals of 667 fs) across the transition at
  metastep 35. The initial diamond structure (a) is strongly strained,
  compressed along one axis and elongated along perpendicular ones (see
  also Fig.1 (a)).  In the next two snapshots (b), (c) the gradual
  disappearance of the diamond structure can be observed; at the same time,
  a new periodic structure emerges (d). (e) - (h) Evolution during 15
  subsequent steps of metadynamics.  Note the gradual formation of the
  simple hexagonal phase.  From the analysis of diffraction peaks the final
  supercell (h) was found to contain 222 primitive cells; therefore 6
  vacancies are actually present in the structure.}
\label{sequence}
\end{figure}

\end{document}